\documentstyle[preprint,aps,epsf]{revtex}

\ifx\epsffile\undefined
\def\insertfig#1#2{}
\else
\def\insertfig#1#2{{\baselineskip=4pt
\centerline{\epsfxsize=\hsize\epsffile{#2}}{{
\centerline{#1}}}\medskip}}
\fi

\def\blmscale{\mu_{\rm BLM}}

\begin{document}
{\tighten
\preprint{\vbox{\hbox{UTPT 94-24}
\hbox{CMU-HEP 94-29}
\hbox{CALT-68-1950}
\hbox{DOE-ER/40682-83}
}}

\title{Perturbative Strong Interaction Corrections to the Heavy
Quark Semileptonic Decay Rate}
\author{Michael~Luke}
\address{Department of Physics, University of Toronto, Toronto, Ontario,
Canada M5S 1A7}
\author{Martin J.~Savage}
\address{Department of Physics, Carnegie Mellon University, Pittsburgh PA
15213}
\author{Mark B.~Wise}
\address{Department of Physics, California Institute
of Technology, Pasadena, CA 91125}

\bigskip
\date{Revised version, September 1994}
\maketitle
\begin{abstract}
We calculate the part of the order $\alpha_s^2$  correction to the
semileptonic heavy  quark decay rate proportional to the number of light
quark flavors, and use our result to set the  scale for evaluating the
strong coupling in the order $\alpha_s$ term according to the  scheme of
Brodsky, Lepage and Mackenzie.  Expressing the decay rate in terms of
the heavy quark pole mass $m_Q$, we find the scale for the
$\overline{MS}$ strong coupling to be $0.07\, m_Q$.   If the decay
rate is expressed in terms of the
$\overline{MS}$ heavy quark mass $\overline m_Q(m_Q)$ then the
scale is $0.12\, m_Q$.  We use these results along with the existing
calculations for hadronic $\tau$ decay to calculate the BLM scale for
the nonleptonic decay width and the semileptonic branching ratio.
The implications for the value of
$|V_{bc}|$ extracted from the  inclusive semileptonic $B$ meson decay
rate are discussed.
\end{abstract}

}

\pacs{}

Inclusive semileptonic $B$ decay has received considerable attention
both theoretically and experimentally.  In the limit where the
$b$ quark mass is much larger than the QCD scale the $B$ meson decay rate
is equal to the $b$ quark decay rate \cite{cgg}.
Corrections to this first arise at order
$(\Lambda_{QCD}/m_b)^2$ and these nonperturbative
corrections may be written in terms of the matrix elements
\cite{bsuv,mb,man}
$\langle B| \bar b (iD)^2 b|B\rangle$ and
$\langle B| \bar b ig \sigma_{\mu\nu} G^{\mu\nu} b|B\rangle$.   The
measured semileptonic $B$ decay rate provides a method  for
determining the magnitude of the element of the
Cabibbo--Kobayashi--Maskawa  matrix $V_{cb}$.  To one loop,
the $b$ quark decay rate is
\begin{eqnarray}\label{bdecayrate}
\Gamma (b \rightarrow X_c e\bar\nu_e) &=& |V_{cb}|^2\  {G_F^2
m_b^5 \over 192\pi^3} \  f (m_c/m_b)\times \nonumber \\
&&\left[1 -{2\alpha_s \over 3\pi} \left(\pi^2 - {25\over 4}
+\delta_1 (m_c/m_b)\right) +...\right].
\end{eqnarray}
In Eq.~(\ref{bdecayrate}) $m_b$ and $m_c$ are the pole masses of the
$b$ and $c$ quarks, $f(x)$ is defined by
\begin{equation}
f(x) = (1 - x^4) (1 - 8 x^2 + x^4) - 24 x^4 \ln x
\end{equation}
and $\delta_1$ takes into
account the effects of the charm quark mass on the order $\alpha_s$
contribution to  the $b$ quark decay rate \cite{jk}. For $m_c/m_b = 0.3$,
$\delta_1=-1.11$.

In Eq.~(\ref{bdecayrate}) the scale of the strong coupling $\alpha_s$ is
usually taken to be $\sim m_b$.  The size of the order
$\alpha_s$ correction depends critically on this choice.  If all of the
higher order terms in the $\alpha_s$ expansion were known then the decay
rate would be independent of the  choice of scale.  However, some choices
of scale give  perturbation series that are badly behaved with higher
orders in the coupling  being very important.  Brodsky, Lepage and
Mackenzie (BLM) \cite{blm}
have  advocated choosing the scale so that
vacuum polarization effects are absorbed  into the running coupling.  This
physically appealing choice of scale usually  results in a reasonable
perturbation series.  In this letter we use our calculation  of the part
of the $\alpha_s^2$ correction proportional to $n_f$ to determine the  BLM
scale appropriate for semileptonic heavy quark decay.

Smith and Voloshin \cite{sv}
have recently shown that the $n_f$
dependent  part of the order $\alpha_s^2$ contribution to the semileptonic
decay rate for a  heavy quark may be written in terms of the one loop
corrections evaluated with a finite gluon mass:
\begin{equation}\label{svint}
\delta\Gamma^{(2)} = {n_f \alpha_s^{(V)}(m_Q)\over 6\pi}  \int_0^\infty
\left(\Gamma^{(1)}
(\mu) - {m_Q^2\over (\mu^2 + m_Q^2)} \Gamma^{(1)} (0)\right)
{d\mu^2\over \mu^2}
\end{equation}
where $\Gamma^{(1)} (\mu)$ is the order $\alpha_s$ contribution
to the decay rate computed with a gluon of mass $\mu$,
and $\alpha_s^{(V)}(m_Q)$ is the strong coupling evaluated in the
$V$-scheme of Brodsky, Lepage and Mackenzie.  This is related to
the usual $\overline{MS}$ coupling $\bar\alpha_s(m_b)$ by \cite{blm}
\begin{equation}
\alpha_s^{(V)} (\mu) = \bar\alpha_s (\mu) + {5\over 3} {\bar\alpha_s^2
(\mu)\over 4\pi}
\left(11 - {2\over 3} n_f\right)\ +... .
\end{equation}
An expression analogous to Eq.~(\ref{svint}) also holds for the
differential rate $d\Gamma/dt$, where $t=(p_e+p_{\bar \nu})^2$.
We have calculated
$\delta\Gamma^{(2)}$ for semileptonic
$Q \rightarrow X_q e\bar\nu_e$ decay with a massless quark $q$ in the final
state and $m_Q  \ll m_W$. The contribution of the graphs
containing a virtual gluon loop to the differential rate
$d\Gamma/dt$ with a massive gluon was calculated analytically
while the integral over the $c$ quark energy in the bremmstrahlung graphs
was performed numerically. The infrared divergences were shown explicitly
to cancel in the sum, and the final integral over the gluon mass was
performed numerically.  Finally, the $t$ integral was also performed
numerically to obtain the correction to the total rate.
In the $t\rightarrow 0$ limit we reproduce the results
given in \cite{sv} for top quark decay in the limit $m_t\gg m_W$.

We find for the total decay width
\begin{eqnarray}\label{twoloop}
\Gamma (Q \rightarrow X_q e\bar\nu_e) = |V_{Qq}|^2\ {G_F^2 m_Q^5\over 192
\pi^3}
&&\left\{1 - {2\alpha_s^{(V)} (m_Q)\over 3\pi} \left(\pi^2 - {25\over
4}\right)\right.\nonumber \\
&&      +\left.  {2\over 3} n_f \left({\alpha_s^{(V)} (m_Q)\over
\pi}\right)^2\ [2.22]+\ \ ...\right\}.
\end{eqnarray}
In terms of the $\overline{MS}$ coupling Eq.~(\ref{twoloop}) becomes
\begin{eqnarray}\label{deltwob}
\Gamma (Q \rightarrow X_q e \bar\nu_e) =  |V_{Qq}|^2\  {G_F^2 m_Q^5\over
192\pi^3}
&& \left\{1 - {2\bar\alpha_s (m_Q)\over 3\pi}  \left(\pi^2 - {25\over
4}\right)\right. \nonumber \\
&& + \left.{2\over 3} n_f  \left({\bar\alpha_s (m_Q)\over
\pi}\right)^2 \ [3.22]\ +\ ...\right\}.
\end{eqnarray}

In Eq.~(\ref{deltwob}) the term proportional to
$n_f$  can be absorbed into the order $\bar\alpha_s$ term if the scale is
changed  from $m_Q$ to $\blmscale$, where
\begin{equation}\label{scalea}
\blmscale = m_Q\ \exp \left\{\ {-3\over (\pi^2 - 25/4)}\ [3.22]\ \right\}
\simeq 0.07 \, m_Q.
\end{equation}
The BLM scale for inclusive heavy quark decay is therefore significantly
smaller than the na\"\i ve estimate of $m_Q$.  In Fig.~1 we plot
the BLM scale for the differential rate $d\Gamma/dt$ as a function
of the squared invariant mass $t$ of the lepton pair.  At
$t=0$ we find the scale $\blmscale=0.12\,m_Q$, which coincides with the
BLM scale found in \cite{sv} for
top quark decay in the limit $m_W\ll m_t$.  As would be expected on physical
grounds, $\blmscale$ decreases as the invariant mass of the
lepton pair increases.

The expression for the width found in Eq.~(\ref{deltwob}) is given in terms
of the pole  mass $m_Q$ of the heavy quark.  The BLM scale $\blmscale$ is
different from that found in Eq.~(\ref{scalea}) if the rate is expressed in
terms of the running
$\overline{MS}$  heavy quark mass evaluated at $m_Q$.
Using \cite{gbgs}
\begin{equation}\label{massa}
m_Q = \bar m_Q (m_Q) \left\{1 \ +\  {4\over 3} {\bar\alpha_s (m_Q)
\over\pi} +(16.11 - 1.04 n_f)\left({\bar\alpha_s
(m_Q)\over\pi}\right)^2 \ +\ ...\right\}
\end{equation}
the semileptonic decay rate becomes
\begin{eqnarray}\label{semifinal}
\Gamma (Q \rightarrow X_q e\bar\nu_e) =
|V_{Qq}|^2\ {G_F^2 [\bar m_Q (m_Q)]^5\over 192\pi^3} &&
\left\{1 - {2\bar\alpha_s (m_Q)\over 3\pi}\  \left(\pi^2 - {65\over
4}\right)\right. \nonumber \\
&&      +\left.  {2\over 3} n_f \left( {\bar\alpha_s (m_Q)\over
\pi} \right)^2  [-4.58]
\ +\ ...\right\}.
\end{eqnarray}
Now the scale $\blmscale$ for which vacuum polarization effects are
absorbed into
the strong coupling is
\begin{equation}\label{blmmsbar}
\blmscale = m_Q\ \exp\left\{\ {-3\over (\pi^2 - 65/4)}\  [-4.58]\ \right\}
\simeq 0.12\, m_Q.
\end{equation}

It has been argued \cite{sv} that a low BLM scale, indicating large
two-loop corrections when $\alpha_s(m_Q)$ is used as an expansion
parameter, would
be expected when relating a ``long-distance" quantity such as the heavy
quark pole
mass to the ``short-distance'' decay rate.  However, our results show
that even if the ``short-distance''
$\overline{MS}$ heavy quark mass is used, the BLM scale $\blmscale$ for the
order $\bar\alpha_s$  correction to semileptonic heavy quark decay
is still significantly less than $m_Q$.  For $b$ decay (neglecting the charm
quark mass) the scale is about 500 MeV and for $c$ decay rate it is only
about 150 MeV. These low scales suggest that QCD
perturbation theory cannot be used for inclusive semileptonic $D$ or
$\Lambda_c$ decay and that an accurate extraction of $|V_{cb}|$ from  the
inclusive semileptonic $B$ decay rate \cite{vbcref}
is not possible without
including all terms of order $\bar\alpha_s^2 (m_Q)$ (and perhaps even
higher orders in $\bar\alpha_s$) in the theoretical expression for the
semileptonic decay rate.

We also note that the BLM scale for inclusive semileptonic heavy quark
decay is somewhat smaller (relative to the heavy quark mass) than the
analogous scale for hadronic
$\tau$ decay.  From the two-loop expression for the inclusive $\tau$ width
\cite{tauref},
\begin{equation}\label{tauwidth}
{\Gamma(\tau\rightarrow\nu_\tau+{\rm hadrons})\over
\Gamma(\tau\rightarrow\nu_\tau\bar\nu_e e^-)}=3\left(1+{\bar\alpha_s(m_\tau)
\over\pi}+(6.340-0.379 n_f)\left({\bar\alpha_s(m_\tau)\over\pi}\right)^2+
...\right)
\end{equation}
the BLM scale for the one-loop expression is found to be $\exp(-3\times
0.379)m_\tau= 0.32\, m_\tau$.   Therefore, although inclusive $\tau$ and
$c$ decays involve comparable energy scales, perturbative QCD is likely to be
at best applicable only to the former.

It is straightforward to extend our results to the case
of nonleptonic heavy quark decay to massless products if we neglect
the
running of the Hamiltonian between $m_W$ and $\mu$.  At order
$\alpha_s$, the corrections to the nonleptonic width are given by two
classes of
diagrams: those with gluons dressing the
$\bar c b$ vertex and those with gluons dressing the $\bar d u$ vertex.
The first
class is identical to that encountered in semileptonic decay, while
the second gives the corrections to $\tau$ decay.  Combining
Eqs.~(\ref{tauwidth}) and (\ref{deltwob}) and including the appropriate colour
factors, we find the expression for the total nonleptonic width to massless
quarks
\begin{eqnarray}\label{nlexpression}
\Gamma_{nl}\equiv\Gamma (Q \rightarrow X_{q_1} X_{q_2} X_{\bar q_3} ) =
|V_{Qq_1}|^2 |V_{q_2 q_3}|^2\  {G_F^2 m_Q^5\over 64\pi^3} && \left\{1 -
{2\bar\alpha_s (m_Q)\over 3\pi}  \left(\pi^2 - {31\over 4}\right)\right.
\nonumber
\\   && +
\left.{2\over 3} n_f
\left({\bar\alpha_s (m_Q)\over
\pi}\right)^2 \ [2.65]\ +\ ...\right\}.
\end{eqnarray} Because the one-loop correction to $\Gamma_{nl}$ is much smaller
than for the semileptonic width $\Gamma_{sl}$, while the order $n_f \alpha_s^2$
terms are comparable, this results in a very low BLM scale for the nonleptonic
width:
\begin{equation}\label{scalenl}
\blmscale = m_Q\ \exp \left\{\ {-3\over (\pi^2 - 31/4)}\ [2.65]\ \right\}
\simeq 0.02 \, m_Q.
\end{equation}
Such a low scale should not be taken literally.  It simply indicates
that the two-loop corrections to
$\Gamma_{nl}$ are significant, requiring an extremely low scale for
$\alpha_s$ in the one-loop term to absorb the order $n_f\alpha_s^2$ terms.

We may also set the BLM scale for the semileptonic branching fraction for
decays
to massless fermions\cite{semilep}.   The contribution from the
class of graphs dressing the
$\bar b c$ vertex cancels in the ratio $\Gamma_{sl}/(\Gamma_{nl}+\Gamma_{sl})$,
and
the corrections are given solely by the graphs which contribute to $\tau$
decay{\footnote{This was also recently noted in \cite{bpbg}.}}.
We thus find, independent of the our results for $\Gamma_{sl}$,
\begin{equation}\label{slfraction} {\Gamma_{sl}\over
\Gamma_{nl}+\Gamma_{sl}}={1\over 4}\left(1-{3\over 4}\left[{\alpha_s(m_Q)\over
\pi} -  0.379 n_f
\left({\alpha_s(m_Q)\over\pi}\right)^2+ ...\right]\right)
\end{equation}
(taking $|V_{q_2 q_3}|=1$), which gives the same BLM scale relative to $m_Q$
as in $\tau$ decay: $\mu_{BLM}=0.32\,m_Q$.

\begin{acknowledgements}

This work was supported in part by the U.~S.~ Department of Energy
under grants DE-FG03-92-ER40701 and DE-FG02-91ER40682, and by the
Natural Sciences and Engineering Research Council of Canada.

\end{acknowledgements}

\vfill\eject
\centerline{\bf Figure Captions}

\begin{enumerate}

\item[1. ] The BLM scale for the partial width $d\Gamma/dt$ (in terms of
the pole mass $m_Q$) as a
function of the lepton pair invariant mass squared $t$.

\end{enumerate}
\vskip 1in
\insertfig{Figure 1}{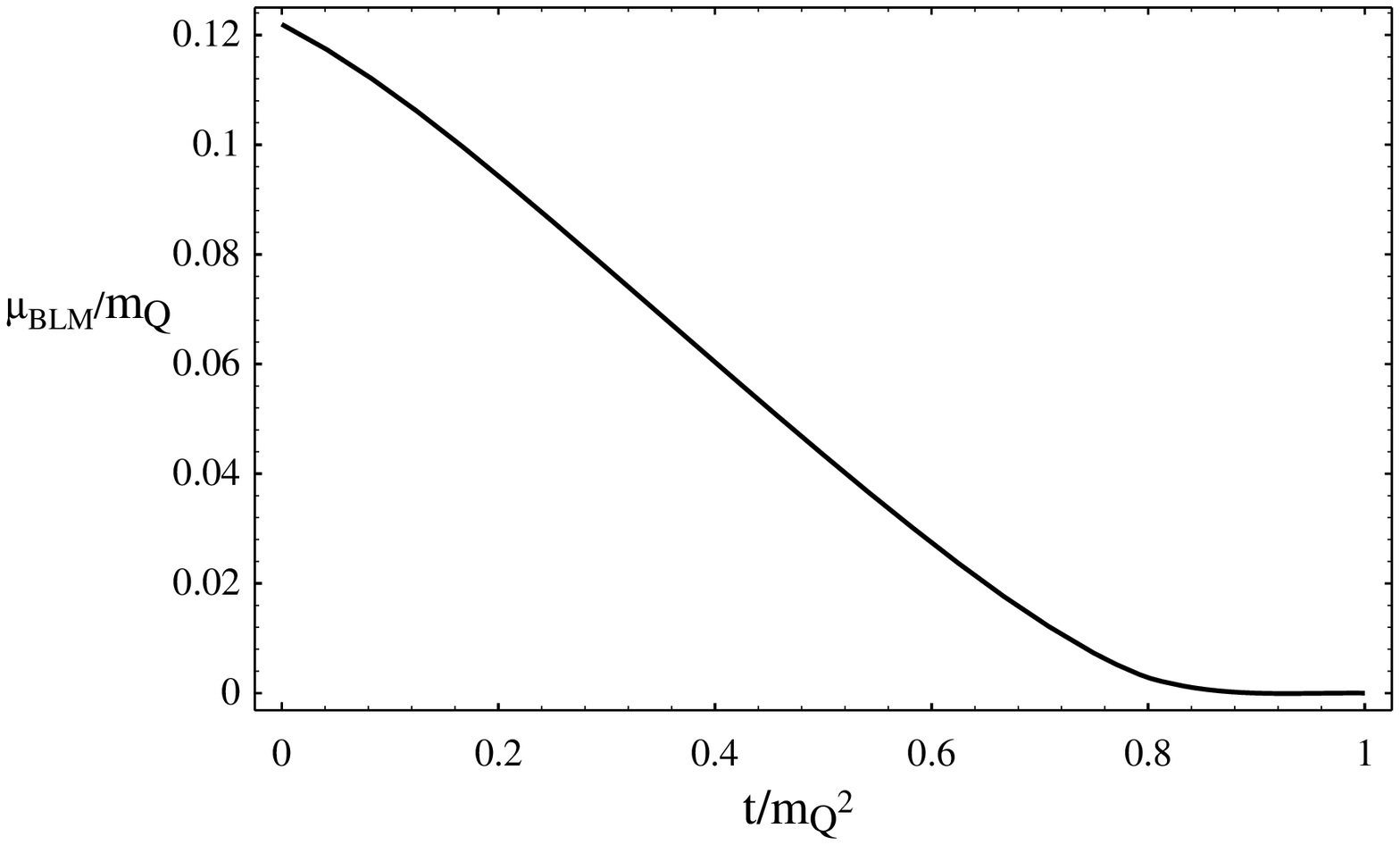}

\end{document}